 \documentstyle[aps]{revtex}   

\newcommand\bom{{\mbox{\boldmath $\omega$}}}

\begin{document}
\draft

\title{Polarization of superfluid turbulence}

\author{Carlo F. Barenghi, Sarah Hulton and David C. Samuels}
\address{Mathematics Department, University of Newcastle,
Newcastle NE1 7RU, UK}
\date{\today}
\maketitle

\begin{abstract}
We show that normal fluid eddies in turbulent helium~II polarize the tangle of
quantized vortex lines present in the flow, thus inducing superfluid
vorticity patterns similar to the driving normal fluid eddies.
We also show that the polarization is effective
over the entire inertial range. The results help explain the
surprising analogies between classical and superfluid turbulence 
which have been observed recently.
\end{abstract}

Our concern is the experimental evidence that
turbulent helium~II
appears similar to classical hydrodynamic 
turbulence\cite{BDV2}. 
For example, Smith et al\cite{grid}
found that the temporal decay of helium II turbulence behind a towed grid 
is the same as that expected in an
ordinary fluid. Maurer \& Tabeling\cite{tabeling} observed the
classical Kolmogorov $k^{-5/3}$ dependence of the energy spectrum
on the wavenumber $k$ in helium II continuously
agitated by rotating propellers. In other experiments it was found
that if helium~II is forced at high velocity along pipes and channels 
the same pressure drops\cite{vansciver-pipe} are detected which are
observed in a classical liquid. Furthermore, when a sphere moves at high
velocity in helium~II, the same drag crisis\cite{vansciver-sphere}
is measured that occurs if the fluid is air.

These apparently classical results are surprising because helium~II is
a quantum fluid. According to Landau's two-fluid theory, it
consists of the intimate mixture of an inviscid superfluid component
and a viscous normal fluid component.  The latter is similar to
a classical Navier - Stokes fluid, so, when made turbulent, it consists
of eddies of various sizes and strengths. On the contrary
quantum mechanics constrains the rotational flow of the
superfluid to discrete quantized vortex lines, each vortex
characterized by the same quantum of circulation 
$\Gamma=9.97~\times 10^{-4}~cm^2/sec$. 
Unlike what happens in a classical Euler fluid, superfluid vortex lines 
can reconnect with each other\cite{recon}. They also interact with
the normal fluid via a linear mutual friction force\cite{BDV1}, and,
when helium~II is made turbulent, they form a disordered, apparently
random tangle.
Clearly the experimental results described above call for an explanation
in terms of the basic physical ingredients of the problem (normal fluid
eddies and superfluid vortices) and their interaction.
What is  remarkable is that these classical
aspects of helium II turbulence are observed to be independent
of temperature, whereas the relative proportion of superfluid and normal
fluid is a strong function of temperature.

A possible solution of the puzzle is that both fluids independently obey the
classical Kolmogorov scaling. A few years ago Nore et al\cite{nore}
solved the Gross-Pitaevskii equation for a tangled system of vortices 
and noticed that the energy spectrum follows a $k^{-5/3}$ dependence.  
Unfortunately the effect only took place over a small range in $k$ space 
and further calculations performed using the vortex filament 
model\cite{tsubota} did not reproduce it.
Another difficulty with this argument is that one has also to 
explain why the mutual friction coupling between the two fluids
does not change the shape of the spectra which the two fluids would
have independently of each other.

A second explanation of the experiments has been suggested
by Vinen\cite{vinen}.  He argued that at spatial scales
larger than the average separation $\delta$ between the quantized vortex
lines, the superfluid and normal fluid are coupled
by a small degree of polarization of the almost random tangle of 
superfluid vortex lines. If that is the case, on these scales
helium~II behaves as a single fluid 
of density $\rho=\rho_s+\rho_n$, consistently with the experiments
($\rho_s$ and $\rho_n$ are the superfluid and normal fluid density
respectively).

The aim of this letter is to support Vinen's argument with quantitative
evidence of polarization. First we shall introduce some simple models
which, although very idealized, capture the essential physics of
polarization. Secondly we shall look for evidence of polarization
by direct numerical simulation. 

Our first model is concerned
with the reaction of superfluid vortices to a normal fluid shear. 
Consider a row of  point vortices of alternating
circulation $\pm \Gamma$ initially set
along the $x$ axis at distance $\delta$ from each other. We assume 
that the normal fluid is ${\bf v}_n=V_n~cos(ky){\bf \hat x}$.
The governing equation of motion of a vortex point is\cite{self-ref}

\begin{equation}
\frac{dy}{dt}=\pm\alpha V_n cos(ky),
\label{equation1}
\end{equation}

\noindent
where $\alpha$ is a known temperature dependent friction
coefficient\cite{BDV1}. The solution of Eq.~(\ref{equation1})
corresponding to the initial condition $y(0)=0$ is

\begin{equation}
y(t)=\frac{2}{k}
\Bigl(
-\frac{\pi}{4}+ tan^{-1}(e^{\pm \alpha k V_n t})\Bigr).
\label{solution1}
\end{equation}

Given enough time ($t\to \infty$), positive and negative vortices
will reach stable locations $y_{\infty}=\pm \pi/2k$ respectively.
In a turbulent normal flow, however, the shear does not last
longer than few times the turnover time
$\tau \approx 1/\omega_n \approx 1/kV_n$. Since $\alpha$ is small
(it ranges from $0.037$ at $T=1.3\rm K$ to $0.35$ at $T=2.16\rm K$)
we have $y(\tau) \approx \pm b$ where $b=\alpha/k$.
Within the lifetime of the shear we have thus created a separation $2b$
between positive and negative vortices, that is to say we have
polarized the initial configuration.
The velocity of this Karman - vortex street is
approximately\cite{karman}
$V_s \approx \pi \Gamma/2 \delta$ for $ k\delta << 1$
in the direction along the $x$ axis where the normal fluid 
(which induced the polarization in the first place) is stronger. 
The result suggests that it is not
necessary to create extra vortex lines to generate a superfluid pattern that
mimics the normal fluid one - rearranging existing vortices is
enough.  We also notice that the induced polarization is proportional
to $\alpha$.

Our second model is concerned with the expansion of favourably
oriented superfluid vorticity. We think of the superfluid vortex tangle as a
collection of vortex rings of radius approximately
equal to the average separation of vortices in the tangle,
$R_0\approx \delta$.  We assume for simplicity that the rings
are on the $x,y$ plane with equal numbers of rings oriented in the $\pm z$
directions.  Depending on whether they have positive or negative
orientation, the rings move along $\pm z$ 
with self induced speed given by

\begin{equation}
V_{R_0}=\frac{\Gamma {\cal L}}{4 \pi R_0},
\label{vel-ring}
\end{equation}

\noindent
where ${\cal L}=\ln{(8R_0/a_0)}-1/2$ is a slowly varying term
and $a_0\approx 10^{-8}\rm cm$ is the vortex core radius. Now we
apply a normal fluid velocity $V_n$ in the $z$ direction.
The radius $R$ of a ring is determined by \cite{BDV1} 

\begin{equation}
\frac{dR}{dt}=\frac{\gamma}{\rho_s \Gamma} (V_n-V_R),
\label{rad-ring}
\end{equation}

\noindent
where $\gamma$ is a known friction
coefficient and $\gamma/\rho_s \Gamma \approx \alpha$ at almost
all temperatures of interest. Eq.~(\ref{rad-ring}) shows that
$V_n$ selectively changes radius and velocity of vortex rings
moving in opposite directions.  A ring which grows (shrinks) by an amount 
$\delta R=  \alpha \delta t (V_n-V_{R_0})$ in time $\delta t$ slows down
(speeds up) by an amount $\delta V_R= V_{R_0}\delta R/R_0$. In this way
a superflow is induced in the same direction of the normal fluid which
induced the polarization in the first place. A simple estimate of the
spatial averaged magnitude of this superflow 
yields\cite{estimate} $V_s\approx3V_{R_0}\delta R/R_0$.

Our third model is concerned with the rotation of existing superfluid
vorticity, We represent a superfluid vortex line as a straight
segment pointing away from the origin
and study how its orientation is changed by a normal fluid
rotation about the $z$ axis. Using spherical coordinates $(r,\theta,\phi)$,
we assume that the vortex is initially in the plane $\theta=\pi/2$.
The normal fluid's velocity is ${\bf v}_n=(0,0,\Omega r\sin{\theta})$
and the motion of the vortex segment is determined by\cite{self-ref2}

\begin{equation}
\frac{d \theta}{dt}=-\alpha \Omega~sin(\theta),
\label{eq-segment}
\end{equation}

\noindent
and $dr/dt=0$  and $d\phi/dt=0$. The solution to
Eq.(\ref{eq-segment}) is

\begin{equation}
\theta(t)=2~tan^{-1}(e^{-\alpha \Omega t}),
\label{sol-segment}
\end{equation}

\noindent
with $r$ and $\phi$ constant.  Given enough time, the vortex segment
will align along the direction of the normal fluid rotation
($\theta \to 0$ for $t \to \infty$), but the lifetime $\tau$ of the eddy
is only of the order of $\tau \approx 1/\Omega$, so the vortex can only turn 
to the angle 

\begin{equation}
\theta(\tau) \approx \pi /2 -\alpha.
\label{angle}
\end{equation}

Despite the smallness of the angle, the effect is sufficient to create a 
net polarization of the tangle
in the direction of the normal fluid's rotation,
provided that there are enough vortices.  The following argument shows how
this is possible. The normal fluid is like a classical viscous
Navier - Stokes fluid, and, if left
to itself, its spectrum $E_k$ would obey Kolmogorov's law

\begin{equation}
E_k=C \epsilon^{2/3} k^{-5/3},
\label{kolmo}
\end{equation}

\noindent
Eq.~(\ref{kolmo}) is valid in the inertial range $1/{\ell}_0 < k < 1/\eta$
in which big eddies break up into smaller eddies, transferring energy to
higher and higher wavenumbers without viscosity playing a role.
Here $k$ is the magnitude of the three dimensional wavevector, $\epsilon$
is the rate of energy dissipation per unit mass, $C$ is a constant of
order unity, ${\ell}_0$ is the integral length scale (the scale at which
energy is fed into the energy cascade) and $\eta$ is the 
Kolmogorov scale at which kinetic energy is dissipated by the
action of viscosity. In reality the normal fluid is not alone but
is forced by the quantized vortex filaments. We know little of the effects
of this forcing
(it has been studied only for very simple geometries\cite{triple})
so, for lack of further information, we assume that the classical
relation Eq.~(\ref{kolmo}) is valid for the normal fluid.

The quantity which is often used to describe the intensity of the superfluid
vortex tangle is the vortex line density $L$ (length $\Lambda$ of vortex 
line per volume ${\cal V}$) because it is easily measured by detecting 
the attenuation of second sound. From $L$ one infers the average 
separation between vortices, $\delta \approx L^{-1/2}$. The quantity
$\Gamma L$ can be interpreted as the total rms vorticity of the superfluid. 
Note that the net amount of superfluid vorticity
in a particular direction can be much less than $\Gamma L$ (even zero, if the 
tangle is randomly oriented).

The key question is whether, as a result of mutual friction, sufficient
quantized vortex lines can re-orient themselves within a normal fluid 
eddy of wavenumber $k$ so that the resulting net superfluid vorticity 
matches the vorticity $\omega_k$ of that eddy.  The process must take place 
in a time scale shorter than the typical lifetime
of the eddy, which is of the order of few times the turnover
time $1/\omega_k$. At this point we use the result of the third model,
for which the initial condition $\theta(0)=\pi/2$ represents the
average case.  In doing so we remark that, if the initial orientation 
of the vortex is toward the origin rather than away from it,
then the vortex segment turns to $\theta(\tau) \approx \pi/2+\alpha$
rather than $\pi/2-\alpha$
but still contributes to positive vorticity in the $z$ direction.
Therefore in the time $1/\omega_k$, re-ordering of existing
vortex lines creates a net superfluid vorticity $\omega_s$ of the order
of $\alpha L \Gamma/3$ in the direction of the vorticity $\omega_k$
of the driving normal fluid eddy of wavenumber $k$.
Since $\omega_k$ is approximately 
$\omega_k=\sqrt{k^3 E_k}$ we have
$ \omega_k=C^{1/2} \epsilon^{1/3} k^{2/3}$. Matching $\omega_s$ and
$\omega_k$ would then require

\begin{equation}
\frac{1}{3} \alpha \Gamma \rm L \ge C^{1/2} \epsilon^{1/3} k^{2/3}.
\label{matching}
\end{equation}

\noindent
The normal fluid vorticity increases with $k$ and is concentrated
at the smallest scale ($k \approx 1/\eta$), so a vortex tangle
with a given value of $L$ may satisfy the above equation only up
to a certain critical wavenumber $k_c$. Substituting
$\epsilon = \nu_n^3/\eta^4$
where $\nu_n$ is the normal fluid's kinematic viscosity (the
viscosity of helium II divided by $\rho_n$),
we obtain
$(\delta/\eta)= C^{-1/4}
(\alpha/3)^{1/2}
(\Gamma/\nu_n)^{1/2}
(\eta k_c)^{-1/3}$.
If $k_c \approx 1/\eta$ then 

\begin{equation}
\frac{\delta}{\eta}= C^{-1/4}
\Bigl(\frac{\alpha}{3}\Bigr)^{1/2}
\Bigl( \frac{\Gamma}{\nu_n} \Bigr)^{1/2}
\label{delta-eta-match}
\end{equation}

\noindent
In the temperature range of experimental interest $\Gamma/\nu_n$
ranges from $0.43$ at $T=1.3\rm K$ to $5.86$ at $T=2.15\rm K$,
so $\delta/\eta = {\cal O}(1)$ and we conclude that matching 
of the two vorticities
($k_c \approx 1/\eta$) is possible throughout the inertial range.

Due to the computational cost, numerical simulations of 
superfluid turbulence do not produce vortex tangles dense enough to 
cover the range $k<1/\ell$ and determine
unambiguously the dependence of the energy spectrum 
on $k$ in this range.  To make progress in the problem and confirm
the above arguments we study the reaction of the superfluid vortex tangle to a 
single scale ABC normal flow\cite{ABC} given by
${\bf v}_n=(A~\sin{(kz)}+C~\cos{(ky)},B~\sin{(kx)}+A~\cos{(kz)},
C~\sin{(ky)}+B~\cos{(kx)})$
where $k=2\pi/\lambda$ is the wavenumber, $\lambda$ is the wavelength
and $A$, $B$ and $C$ are parameters.  ABC flows are solutions of the
Euler equation and the forced Navier - Stokes equation and have been used 
as idealized model of eddies in fluid dynamics, magneto-hydrodinamics
and superfluid hydrodynamics\cite{ABC2}.  For the sake of simplicity we take 
$A=B=C$ and $\lambda=1$.

Following Schwarz\cite{schwarz}, we represent a superfluid
vortex filament as a space curve ${\bf s}={\bf s}(\xi,t)$ where $\xi$ is 
arclength and $t$ is time. Neglecting a small transverse friction coefficient,
the curve moves with velocity

\begin{equation}
\frac{d{\bf s}}{dt}
={\bf v}_{si} + \alpha {\bf s}' \times
({\bf v}_n -{\bf v}_{si}).
\label{velocity}
\end{equation}

\noindent
where ${\bf s}'=d{\bf s}/d\xi$ and the self induced velocity
${\bf v}_{si}$  is given by the Biot - Savart integral

\begin{equation}
{\bf v}_{si}=\frac{\Gamma}{4\pi}
 \int {{( {\bf r} - {\bf s} ) \times {\bf dr}}
\over{\vert  {\bf r} - {\bf s} \vert^3}}.
\label{biot-savart}
\end{equation}

\noindent
The calculation is performed in a cubic box of volume 
${\cal V}=1~cm^3$ with periodic boundary conditions.
The numerical technique is standard\cite{schwarz} and the details of our 
algorithm, including how to perform vortex reconnections,
have been published elsewhere\cite{ABC2}. 

We start the calculation with $N=50$ superfluid vortex rings
set at random positions and orientation 
and integrate  in time at a variety of temperatures 
($\alpha=0.1$, $0.5$ and $1.0$) and normal fluid's velocities 
($A=0.01$, $0.1$, $1.0$ and $10.0~cm/sec$). The vortex length 
($\Lambda=76.8~cm$ at $t=0$)
increases or decreases depending on whether the ABC flow is
strong enough to feed energy into the normal fluid via instabilities of 
vortex waves (for example, for $\alpha=1.0$, the final length
$\Lambda$ is as high as $781.3~cm$ at 
$A=10.0~cm/sec$, and as low as $56.67~cm$ at $A=0.01~cm/sec$)
The rings interact with each other and
with the normal fluid, get distorted, reconnect, and soon an apparently
random tangle is formed (see figure 1).

The quantity
$<\cos{(\theta)}>=<{\bf s}' \cdot {\hat \bom_n>}$, which we monitor
during the evolution, gives us
the tangle - averaged projection of the 
local tangent to a vortex in the direction of the local
normal fluid vorticity,
$\hat \bom_n=(1/\omega_n)\bom_n$, where $\bom_n=\nabla \times {\bf v}_n$. 
At $t=0$ $<\cos{(\theta)}>=0$ due to
the random nature of the initial state, and it is apparent from figure 2
that $<\cos{(\theta)}>$ increases with time, no matter whether $\Lambda$
decreases or increases.

The results are analyzed in figure 3.
From the simple models descrived above we expect that the polarization
induced by the normal fluid vorticity is proportional to $\alpha$.
We also know from the discussion above that we should restrict the
analysis to times $t<\tau$ where $\tau=1/\omega_n$ with $\omega_n=\sqrt{3}Ak$
is the lifetime of the normal fluid eddy, which we assume to be the same
as the eddy's turnover time. It is apparent from the figure that, no
matter whether the tangle grows or decays, approximately the
same polarization takes place
for $t/\tau<1$.

In conclusion we have put the theory of superfluid turbulence
on firmer ground. Using simple models which capture the essential
physical mechanisms of polarizaton
and then a numerical simulation, we have shown that,
within the lifetime of a normal fluid eddy of wavenumber $k$, 
superfluid vortex lines can rearrange themselves
so that the superfluid vorticity and the normal fluid vorticity are
aligned. Provided that enough vortex lines are present, vorticity
matching should take place over the entire inertial range, up
to wavenumbers $k$ of the order of $1/\ell$.

Our result has theoretical and experimental implications.
Numerical simulations of vortex lines driven
by normal fluid turbulence\cite{energy} show a $k^{-1}$ superfluid energy
spectrum in the accessible region $k \geq 1/\delta$. More intense
(hence computationally expensive) vortex tangles should be investigated
to explore the region $k<<1/\delta$ where, as a consequence of our
result, we predict the classical $k^{-5/3}$ dependence. Clearly
an important issue which must be investigated is the nonlinear
saturation of the polarization process.
On the experimental side, our result supports the use of helium~II
to study classical turbulence. This has been done recently by Skrbek
et al\cite{SND} who exploited the physical properties of liquid
helium to study the decay of vorticity on an unprecedented wide range
of scales.

\centerline{ACKNOWLEDGMENTS}
\medskip

We are indebted to W.F. Vinen who suggested to study some of the above
described models of the interaction between normal
fluid and quantized vortex lines.

\centerline{FIGURE CAPTIONS}
\medskip

\noindent
{\bf Figure 1}. Vortex configuration at $t=0.123~sec$ for
$\alpha=0.5$ and $A=1.0~cm/sec$.

\noindent
{\bf Figure 2}. Average polarization $<\cos{(\theta)}>$ versus time $t$
computed for different values of $A$ and $\alpha$.

\noindent
{\bf Figure 3}. $<\cos{(\theta)}>/\alpha$ versus scaled time $t/\tau$ where
$\tau$ is the eddy's lifetime (same symbols as in Figure 1).


\begin{references}


\bibitem{BDV2}
C.F. Barenghi, R.J. Donnelly and W.F. Vinen,
{\it Quantized Vortex Dynamics And Superfluid
Turbulence}, Springer Verlag (2001).

\bibitem{grid} 
M.R. Smith, R.J. Donnelly, N. Goldenfeld and W.F. Vinen,
Phys. Rev. Lett. {\bf 71}, 2583 (1993).

\bibitem{tabeling}
J. Maurer and P. Tabeling, Europhys. Lett. {\bf 43}, 29 (1998).

\bibitem{vansciver-pipe}
P.L. Walstrom, J.G. Weisend, J.R. Maddocks and S.W. VanSciver,
Cryogenics {\bf 28}, 101 (1998).

\bibitem{vansciver-sphere}
M.R. Smith, D.K. Hilton and S.V. VanSciver,
Phys. Fluids {\bf 11}, 751 (1999).

\bibitem{recon}
J. Koplik, J. and H. Levine, Phys. Rev. Lett. {\bf 71}, 1375-1378 (1993);
M. Leadbeater, T. Winiecki, D.C. Samuels, C.F. Barenghi
and C.S. Adams, Phys. Rev. Letters {\bf 86} 1410 (2001).

\bibitem{BDV1}
C.F. Barenghi, R.J. Donnelly and W.F. Vinen,
J. Low Temp. Phys. {\bf 52} 189, (1983).

\bibitem{nore}
C. Nore, M. Abid and M.E. Brachet,
Phys. Rev. Lett. {\bf 78} 3896 (1997).

\bibitem{tsubota} 
T. Araki, M. Tsubota and S.K. Nemirowskii,
J. Low Temp. Phys. {\bf 126}, 303 (2002).

\bibitem{vinen}
W.F. Vinen, Phys. Rev. B {\bf 61}, 1410 (2000).

\bibitem{karman}
We estimate the velocity of the double row of vortices at
$(ma,b/2)$ and $((m+1/2)a,-b/2)$ (where $m$ is integer and
$a=2\delta$) at the orgin and obtain
$V_s=(\pi \Gamma/a)sinh(\pi b/a)/[cosh(\pi b/a)-1]
\approx \pi \Gamma/2 \delta$.

\bibitem{estimate}
Let $n$ be the number of positive and negative
rings. Their areas and velocities are respectively $A^{\pm}=\pi R_0^2$ and
$V^{\pm}=\pm V_{R_0}$. At $t=0$ the average velocity is
$V_s=[n \pi R_0^2 V_{R_0}-n \pi R_0^2 V_{R_0}]/[n\pi R_0^2 + n \pi R_0^2]=0$. 
At time $t=\delta t$ we have 
$A^{\pm}=\pi(R_0\pm\delta R)^2$ 
and $V^{\pm}=\pm(V_{R_0} \mp \delta V_R)$, so
$V_s=[n \pi(R_0+\delta R)^2(V_{R_0}-\delta V_R)
-n(R_0-\delta R)(V_{R_0}+\delta V_R)]/
[n\pi(R_0+\delta R)^2+n\pi(R_0-\delta R)^2]\approx 3V_{R_0}\delta R/R_0$.

\bibitem{self-ref}
Eq.~(\ref{equation1}) follows from Eq.~(\ref{velocity}) in the absence
of ${\bf v}_{si}$.

\bibitem{self-ref2}
Again, Eq.~(\ref{eq-segment}) follows from Eq.~(\ref{velocity}) neglecting the
self induced motion which is zero for a straight vortex.

\bibitem{triple}
D. Kivotides, C.F. Barenghi and D.C. Samuels,
Science, {\bf 290}, 777 (2000);
O.C. Idowu, A. Willis, C.F. Barenghi and D.C. Samuels,
Phys. Rev. B {\bf 62} 3409 (2000).

\bibitem{ABC}
T. Dombre, U. Frisch, J.M. Greene, M. Henon, A. Mehr and A.M. Soward,
J. Fluid Mechanics {\bf 167}, 353 (1986).

\bibitem{ABC2}
C.F. Barenghi, D.C. Samuels, G.H. Bauer and R.J. Donnelly,
Phys. Fluids {\bf 9}, 2631 (1997).

\bibitem{schwarz}
K.W. Schwarz, Phys. Rev. Letters {\bf 49}, 283 (1982); 
Phys. Rev. B {\bf 31}, 5782 (1985);
Phys. Rev. B {\bf 38}, 2398 (1988).


\bibitem{energy}
D. Kivotides, J.C. Vassilicos, D.C. Samuels and C.F. Barenghi,
Europhys. Letters, {\bf 57} 845 (2002).

\bibitem{SND}
L. Skrbek, J.J. Niemela and R.J. Donnelly,
Phys. Rev. Lett. {\bf 85}, 2973 (2000).

\end{references}
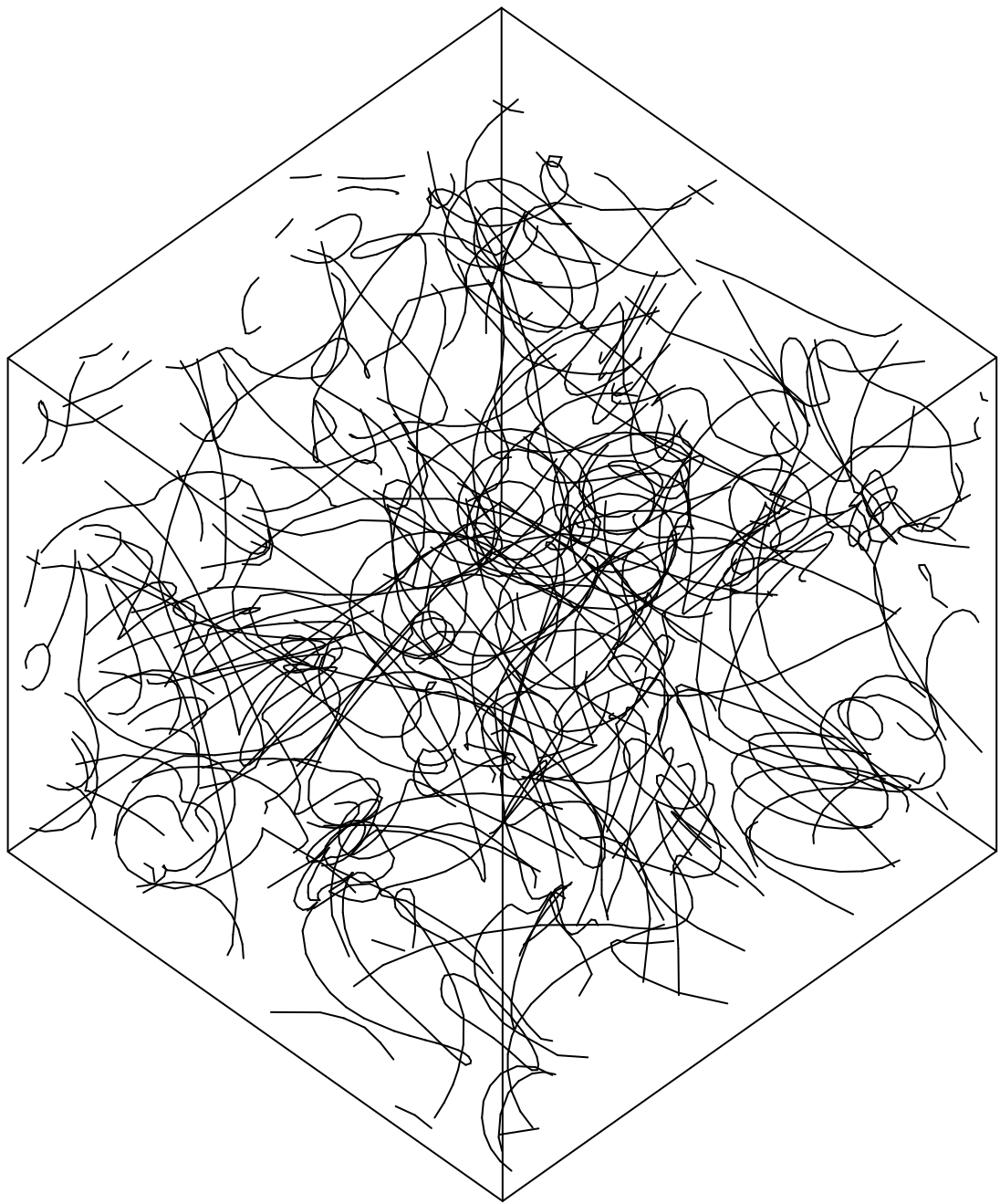
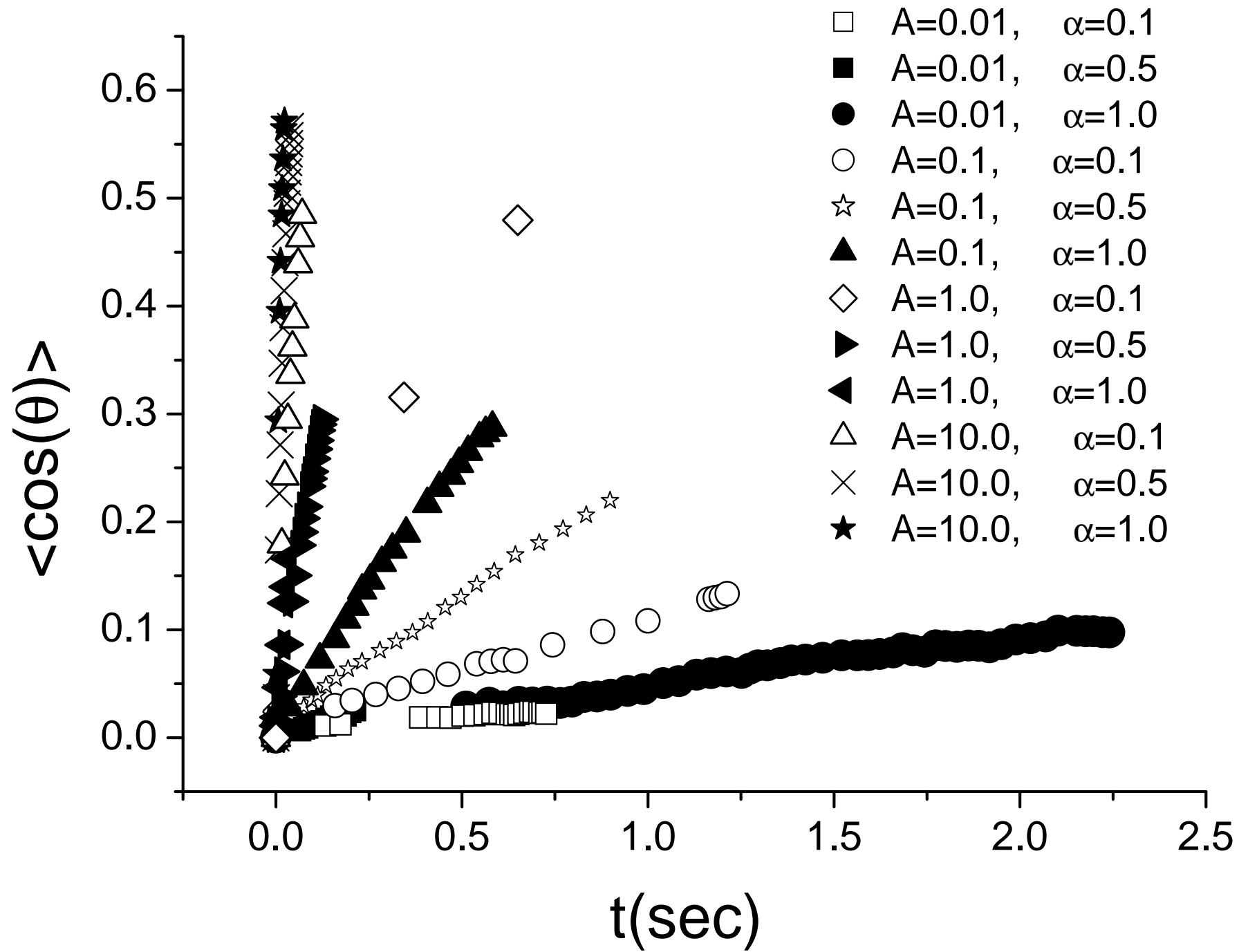
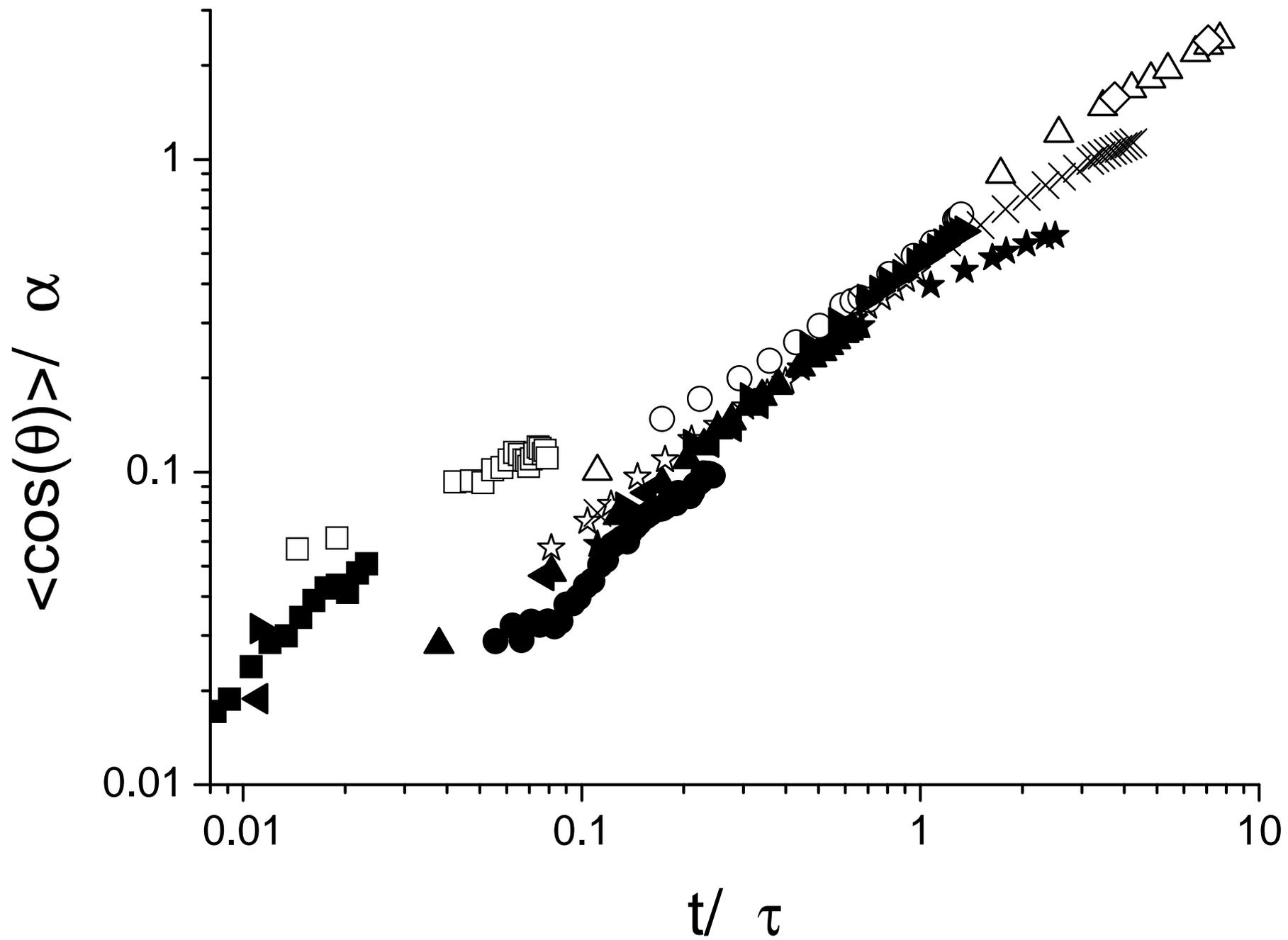
\end{document}